\documentclass[12pt]{article}   
   
\usepackage{array}   
\usepackage{epsfig}   
\usepackage{amssymb}   
\usepackage{graphics,graphpap}   
\usepackage{amssymb}   
\usepackage{amsmath}   
\usepackage[usenames]{color}  
\usepackage{slashed}   
\usepackage{cite}   
\usepackage{hyperref}
\usepackage{graphicx}

\renewcommand{\thefootnote}{\fnsymbol{footnote}}   
   
\def\s#1{\setbox0=\hbox{$#1$}%
  \rlap{\ifdim\wd0>.7em\kern.22\wd0\else\kern.1\wd0\fi /}#1}   


\numberwithin{equation}{section}

\begin{document}   
   
\begin{titlepage}   
\begin{flushright}   
\begin{tabular}{l}   
IPPP-14-100\\DCPT-14-200\\MITP/14-097
\end{tabular}   
\end{flushright}   
\vskip1.5cm   
\begin{center}   
{\Large \bf \boldmath   
New physics effects in tree-level decays 

\vspace{0.1cm}

and 

\vspace{0.3cm}

the precision in the determination of the CKM angle $\gamma$}   
\vskip1.3cm    
{\sc   
Joachim Brod      \footnote{joacbrod@uni-mainz.de}$^{,1}$,
Alexander Lenz    \footnote{Alexander.Lenz@durham.ac.uk}$^{,2}$,
Gilberto Tetlalmatzi-Xolocotzi \footnote{gilberto.tetlalmatzi-xolocotz@durham.ac.uk}$^{,2}$,
Martin Wiebusch   \footnote{Martin.Wiebusch@durham.ac.uk}$^{,2}$,

}   
\vskip0.5cm   
$^1$ Mainz Institute for Theoretical Physics, Johannes Gutenberg University,
     55099 Mainz, Germany 
\\
$^2$ Institute for Particle Physics Phenomenology (IPPP), 
      Department of Physics, Durham University, DH1 3LE, United Kingdom

\vskip2cm

   
   
{\large\bf Abstract\\[10pt]} \parbox[t]{\textwidth}{We critically
  review the assumption that no new physics is acting in tree-level
  $B$-meson decays and study the consequences for the ultimate
  precision in the direct determination of the CKM angle $\gamma$. In
  our exploratory study we find that sizable universal new physics
  contributions, $\Delta C_{1,2}$, to the tree-level Wilson
  coefficients $C_{1,2}$ of the effective Hamiltonian describing weak
  decays of the $b$ quark are currently not excluded by experimental
  data.  In particular we find that Im $\Delta C_1 $ and Im $\Delta
  C_2 $ can easily be of order $\pm 10\%$ without violating any
  constraints from data.  Such a size of new physics effects in $C_1$
  and $C_2$ corresponds to an intrinsic uncertainty in the CKM angle
  $\gamma$ of the order of $|\delta \gamma| \approx 4^\circ$, which is
  slightly below the current experimental precision. The accuracy in
  the determination of $\gamma$ can be improved by putting stronger
  constraints on the tree-level Wilson coefficients, in particular
  $C_1$. To this end we suggest a more refined theoretical study as
  well as a more precise measurements of the observables that
  currently provide the strongest bounds on hypothetical new weak
  phases in $C_1$ and $C_2$. We note that the semi-leptonic CP
  asymmetries seem to have the best prospect for improving the bound
  on the weak phase in $C_1$.}
   
\vfill   
   
\end{center}   
\end{titlepage}   
   
\setcounter{footnote}{0}   
\renewcommand{\thefootnote}{\arabic{footnote}}   
\renewcommand{\theequation}{\arabic{section}.\arabic{equation}}   
   
\newpage   
   
\section{Introduction}   

The standard model of particle physics (SM) seems to be more
successful than previously expected. With the detection of the Higgs
particle in 2012 its particle content is finally complete.  Up to now
we have neither directly detected new particles nor did we find
significant new physics effects in indirect searches. Nevertheless,
many of the motivations for new physics searches, like the origin of
the baryon asymmetry in the universe or the nature of dark matter,
remain unanswered within the SM. In addition, there are several hints
for experimental deviations from SM predictions, e.g. in the quark
flavour sector, see for example~\cite{Lenz:2014nka,
  Altmannshofer:2014rta}. In order to draw any definite conclusions
from these arising hints for new physics, a higher precision is
mandatory both in experiment and theory. In that respect also some
unquestioned prejudices that might only be valid as a crude assumption
have to be revisited. In this letter we reconsider the commonly
accepted supposition that there are no new physics effects in
tree-level decays of heavy quarks and show that there is, purely from
the viewpoint of current data, still plenty of room for deviations
from SM predictions. Bounds on the Wilson coefficients of the SM
current-current operators have been obtained, using a restricted set
of observables, already in~\cite{Bobeth:2014rda, Bobeth:2014rra}. Here
we consider a larger set of observables to constrain the SM
current-current sector. Also, we assume that the new-physics effects
are flavor universal. This will give us a general idea of the size of
the effects; we leave a more detailed analysis for future
work~\cite{next}.

As an interesting application of our results we consider the precision
in the determination of the Cabibbo-Kobayashi-Maskawa (CKM) angle
$\gamma$. This angle can be extracted from tree-level $B \to D K$
decays essentially without hadronic
uncertainties~\cite{Bigi:1981qs}. An important assumption for this
analysis is the absence of weak phases other than $\gamma$ in these
decays. While many different corrections to this assumption have been
studied in the literature (see the discussion in
Section~\ref{sec:gamma}), the absence of new-physics contributions to
the tree-level Wilson coefficients has, to our knowledge, hitherto not
been questioned in this context. We emphasise that from a purely
phenomenological viewpoint we cannot exclude shifts in $\gamma$ of the
order of $\pm4^\circ$ that are clearly not negligible in view of the
expected sensitivity of $1^\circ$ at LHCb and
Belle~II~\cite{CERN-LHCC-2011-001, Abe:2010gxa}. Hence, the statement
that the extraction of $\gamma$ from tree-level decays corresponds to
a pure SM value should be taken with care.

This letter is organised as follows. In Section~\ref{sec:NP} we
collect all bounds on the Wilson coefficients of the current-current
operators, and investigate the implication for the extraction of
$\gamma$ in Section~\ref{sec:gamma}. We summarise our findings in
Section~\ref{sec:conclusion}, where we also point out some strategies on 
how to improve the bounds on new physics effects in tree-level decays.

\section{New physics in tree-level decays}\label{sec:NP}

We start our considerations of the possible size of new physics
effects with the effective Hamiltonian for non-leptonic $b$-quark
decays of the form $b \to u_1 \bar{u}_2 d_1$, where $u_{1,2}$ are
up-type quarks and $d_1$ is a down-type quark:
\begin{eqnarray}\label{eq:Heff}
 {\cal H}_{\rm eff.}^{ \bar{u}_1 u_2 d_1} & = & \frac{G_f}{\sqrt{2}} V_{u_1b} V_{u_2d_1}^* 
 \left[ C_1 Q_1^{ \bar{u}_1 u_2 d_1 } +  C_2 Q_2^{ u_1 \bar{u}_2 d_1 } \right],
\end{eqnarray}
with the colour singlet operators $Q_2$ and the colour rearranged
operators $Q_1$, 
\begin{eqnarray}
 Q_1^{ \bar{u}_1 u_2 d_1} = (\bar{u}_1^\alpha b^\beta)_{V-A} (\bar{d}_1^\beta u_2^\alpha)_{V-A} \; ,
&&
 Q_2^{ \bar{u}_1 u_2 d_1} = (\bar{u}_1^\alpha b^\alpha)_{V-A} (\bar{d}_1^\beta u_2^\beta)_{V-A} \; ,
\end{eqnarray}
where $\alpha$ and $\beta$ are colour indices, and $(\bar q q')_{V-A}$
stands for $\bar q \gamma_\mu (1 - \gamma_5) q'$. In this letter we
consider possible new physics contributions to the Wilson coefficients
$C_1$ and $C_2$, denoted by $\Delta C_1$ and $\Delta C_2$, both of
which can in general be complex.  Thus we have for the full Wilson
coefficients
\begin{eqnarray}
C_1 = C_1^{\rm SM} + \Delta C_1\; , && C_2 = C_2^{\rm SM} + \Delta C_2\; .
\end{eqnarray}

A first step in such a direction has been performed recently
in~\cite{Bobeth:2014rda}, where the effect of new physics
contributions to the decays $b \to c \bar{u} d$, $b \to u \bar{c} d$,
$b \to c \bar{c} d$, and $b \to u \bar{u} d$ on the decay rate
difference of the neutral $B_d$ meson system, $\Delta \Gamma_d$, was
investigated. In \cite{Bobeth:2014rra} new physics contributions to
the tree-level part of the $b \to u \bar{u} s$ decay were considered
as solution to the ``$ \Delta {\cal A}_{\rm CP} $ puzzle'' in $B \to K
\pi$ decays. We will not consider the observables from
\cite{Bobeth:2014rra}, because they are very sensitive to penguin
contributions, whereas we concentrate on tree-dominated
decays. Moreover, our final conclusion would not change with the
inclusion of the $B \to K \pi$ observables.

The motivation of this work is to study the effect of new weak phases
on the extraction of $\gamma$ from tree-level decays. We thus extend
the analysis in~\cite{Bobeth:2014rda} which focused on final states
involving a down quark, by including more $b$-decay channels and thus
more observables. However, we make the simplifying assumption that all
possible non-leptonic $b$-quark decay channels receive the same new
physics contributions. Decay-channel specific new physics
contributions would in general give looser bounds on the individual
new physics contributions to $C_1$ and $C_2$, and will be considered
in a forthcoming publication~\cite{next}.

The following observables are taken over directly
from~\cite{Bobeth:2014rda}:
\begin{itemize}
\item The $b \to c \bar{u} d$-transition is constrained by $B \to D
  \pi$ and $B \to D^{(*)0} h^0$ decays. For the corresponding theory
  expressions QCD factorisation~\cite{Beneke:2000ry} is used.
\item The rare decay $b \to d \gamma$ gives the strongest bound on the
  $b \to c \bar{c} d$-transition, where we use the theoretical
  formulae from \cite{Gambino:2001ew} and \cite{Crivellin:2011ba}.
  This decay gets also restrictions~\cite{Bobeth:2014rda} from the
  direct measurement of the CKM angle $\beta$ in the decay $ B \to J /
  \psi K_S$ and the semi-leptonic asymmetry $a_{sl}^d$ described in
  more detail below.
\item QCD factorisation~\cite{Beneke:2001ev} is used again to
  constrain the $b \to u \bar{u} d$-channel with $B \to \pi\pi, \rho
  \pi, \rho \rho$-decays. As in \cite{Bobeth:2014rda} for the
  $B\rightarrow \pi\pi$ transition two observables are considered: the
  indirect CP asymmetry $S_{\pi\pi}$ and the ratio of hadronic and
  differential semi-leptonic decay rate $R_{\pi^-\pi^0}$.
\end{itemize}
For these observables we use the same formalism and the same
experimental data as described in~\cite{Bobeth:2014rda} and we refer
the interested reader to this paper for details. Next we extend some
of the formulae used already in~\cite{Bobeth:2014rda}.
\begin{itemize}
\item The total lifetime of $b$-hadrons can be compared with the
  experimental measurements. We use the following expression that
  shows the explicit dependence on the Wilson coefficients, see e.g.
  \cite{Lenz:2014jha}:
      \begin{equation}
      \frac{\Gamma_{\rm tot}}{\Gamma_{\rm tot}^{\rm SM}} = 
      \frac{3 |C_1|^2          + 3 |C_2|^2          + 2 {\rm Re} [C_1^* C_2]}
           {3 |C_1^{\rm SM}|^2 + 3 |C_2^{\rm SM}|^2 + 2 {\rm Re} [C_1^{* \rm SM} C_2^{\rm SM}]}
           \; .
      \end{equation}
      For $\Gamma_{\rm tot}^{\rm SM}$ we take the result
      from~\cite{Krinner:2013cja} that includes $\alpha_s$-corrections
      and terms that are subleading in the heavy-quark expansion; the
      experimental value is taken from \cite{Amhis:2012bh}:
      \begin{equation}
      \Gamma_{\rm tot}^{\rm SM} = (3.6  \pm 0.8 ) \cdot 10^{-13} \; {\rm GeV} \; ,  \hspace{1cm}
      \Gamma_{\rm tot}          = (4.20 \pm 0.02) \cdot 10^{-13} \; {\rm GeV}  \; .
      \end{equation}

\item For the channel $b \to c \bar{c} s$ we take constraints from the
  branching ratio $\mathcal{B}(B \rightarrow X_s \gamma)$ into
  account. The bounds for this observable were calculated using the
  NLO expressions given in~\cite{Chetyrkin:1996vx} as well as the NNLO
  SM value quoted in \cite{Misiak:2006zs}, the experimental result
  considered was obtained from~\cite{Amhis:2012bh}.
\end{itemize}

Additional bounds on $C_1$ and $C_2$ can be obtained from the decay
rate difference of the neutral $B_s$-mesons, $\Delta \Gamma_s$, and
the semi-leptonic CP asymmetries, $a_{sl}^s$. These observables have
not been considered in~\cite{Bobeth:2014rda}; they can be extracted
for both neutral $B$-meson systems from the theory expression for
$\Gamma_{12}^q/M_{12}^q$:
\begin{eqnarray}
  a_{sl}^q = {\rm Im} \left(\frac{\Gamma_{12}^q}{M_{12}^q}
  \right) \; ,
  \hspace{1cm}
  \frac{\Delta \Gamma_q}{\Delta M_q} = - {\rm Re} \left(\frac{\Gamma_{12}^q}{M_{12}^q} \right)
  \; .
\end{eqnarray}
Using the results from \cite{Beneke:1998sy,Beneke:2003az,Lenz:2006hd}
we find for the explicit dependence on the NP contributions $\Delta
C_1$ and $\Delta C_2$:
\begin{eqnarray}
      \frac{\Gamma_{12}^d/M_{12}^d}{\Gamma_{12}^{d, \rm SM}/M_{12}^{d, \rm SM}} =  
       1 &&
            - \; (0.23  - 0.047  i) \cdot \Delta C_1 
            + \; (0.76  + 0.25   i) \cdot \Delta C_1^2 
            + \; (1.91  - 0.0029 i) \cdot \Delta C_2
      \nonumber \\
      && 
         + \;(0.084 + 0.14   i) \cdot \Delta C_1 \cdot \Delta C_2 
         + \;(0.93  + 0.0072 i) \cdot \Delta C_2^2
      \; ,
      \\
      \frac{\Gamma_{12}^s/M_{12}^s}{\Gamma_{12}^{s, \rm SM}/M_{12}^{s, \rm SM}} =  
       1 &&
            - \; (0.24  + 0.022  i) \cdot \Delta C_1 
            + \; (0.68  - 0.012  i) \cdot \Delta C_1^2 
            + \; (1.90  + 0.00013 i) \cdot \Delta C_2
      \nonumber \\
      && 
         + \;(0.034 - 0.0068   i) \cdot \Delta C_1 \cdot \Delta C_2 
         + \;(0.93  - 0.00035 i) \cdot \Delta C_2^2
      \; .
      \label{Gamma12overM12}
\end{eqnarray}
We now express the semi-leptonic asymmetry and the decay rate
difference in terms of these ratios as
\begin{eqnarray}
        a_{sl}^q & = & {\rm Im} \left(
        \frac{\Gamma_{12}^q/M_{12}^q}{\Gamma_{12}^{q, \rm SM}/M_{12}^{q, \rm SM}} 
        \cdot \frac{\Gamma_{12}^{q, \rm SM}}{M_{12}^{q, \rm SM}} 
       \right)
        \; ,
        \\
        \Delta \Gamma_q &  = & - {\rm Re} \left(
        \frac{\Gamma_{12}^q/M_{12}^q}{\Gamma_{12}^{q, \rm SM}/M_{12}^{q, \rm SM}} 
        \cdot \frac{\Gamma_{12}^{q, \rm SM}}{M_{12}^{q, \rm SM}} 
       \right) \cdot
        \Delta M_q^{\rm Exp.} \; .
\end{eqnarray}
The SM prediction for $\Gamma_{12}^q/M_{12}^q$ is given
in~\cite{Lenz:2011ti} and reads
\begin{eqnarray}
      \frac{\Gamma_{12}^{d, \rm SM}}{M_{12}^{d, \rm SM}}  = 
     -0.0050  - 0.00045 i 
      \; ,
      &&   
      \frac{\Gamma_{12}^{s, \rm SM}}{M_{12}^{s, \rm SM}}  = 
      -0.0050 + 0.000021 i \; .
\end{eqnarray}    
The experimental value for $\Delta \Gamma_s$ is taken
from~\cite{Aaij:2014zsa}, for the semi-leptonic asymmetries we take
the naive average of the values in~\cite{Abazov:2012hha, Lees:2013sua,
  Aaij:2014nxa, Lees:2014qma, Abazov:2012zz, Aaij:2013gta}, and for
the mass difference we use the HFAG average~\cite{Amhis:2012bh}. We
find
      \begin{eqnarray}
        a_{sl}^d  =   (+2.2 \pm  2.2) \cdot 10^{-3} \; ,
       & &
       \Delta \Gamma_s =  0.0805 \pm 0.0091 \pm 0.0032 \, {\rm ps}^{-1} \; ,
        \\
        a_{sl}^s  =   (-4.8 \pm  4.8) \cdot 10^{-3} \; ,
        & &
        \Delta M_s  =  17.761 \pm 0.022 \, \mbox{ps}^{-1} \; .
      \end{eqnarray}
      We do not use $\Delta \Gamma_d$ since there are currently only loose
      experimental bounds available.

\begin{figure}
\begin{center}
\includegraphics[width =0.45 \textwidth]{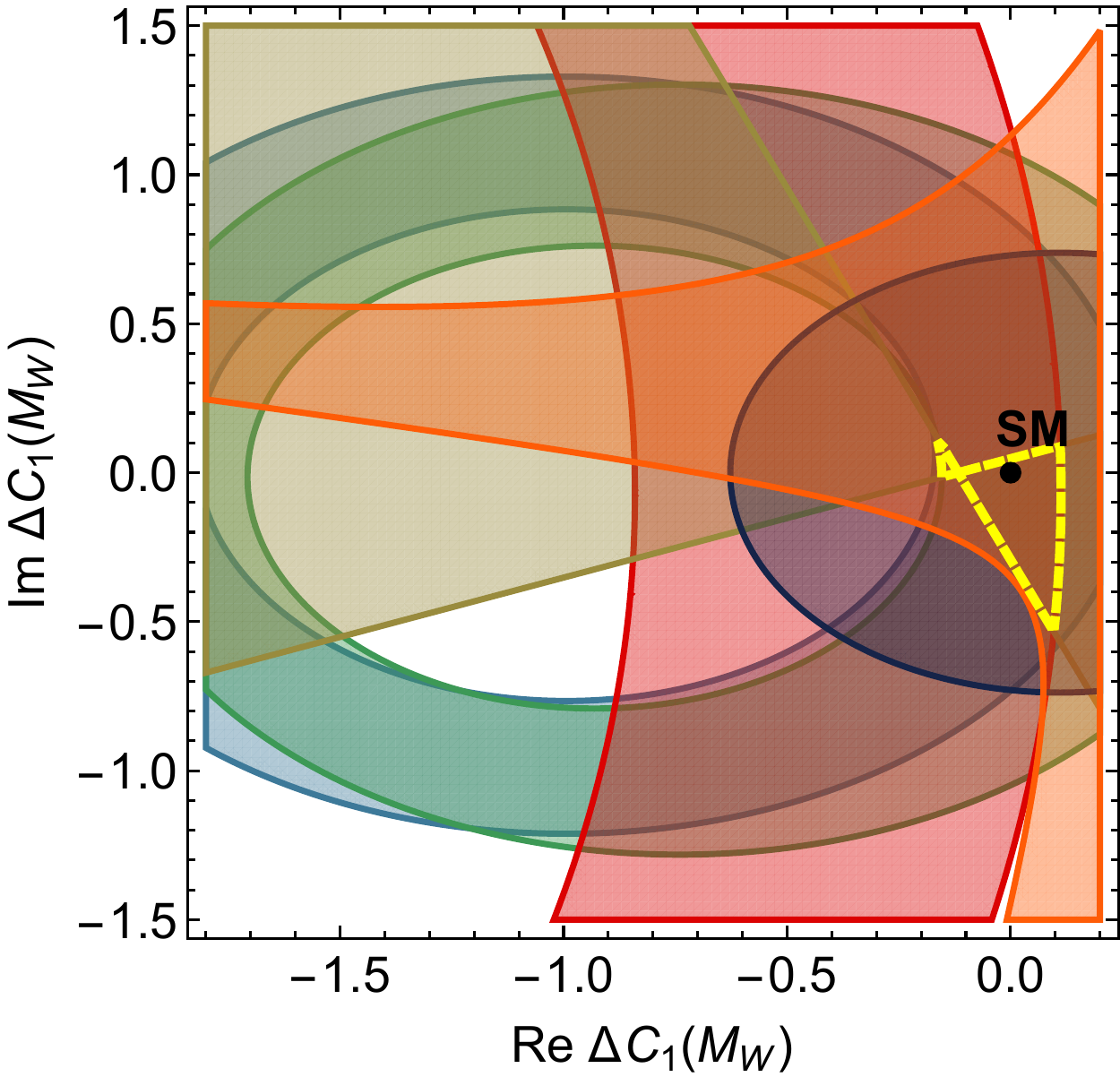}
\end{center}
\caption{Constraints on $\Delta C_1$, the new-physics contribution to
  the tree-level Wilson coefficient $C_1$, at the scale $\mu_W =
  M_W$. The red region is associated with constraints from the $B \to
  D \pi$ decay channel, the green and blue rings with the transitions
  $B \to \rho \rho$ and the observable $R_{\pi^-\pi^0}$ calculated
  from the decay $B \to \pi \pi$, respectively. The brown sections are
  related to the decays $B^0 \to D^{(*)0}h^0$ and the blue circle to
  the total lifetime of $b$-hadrons. Finally, the region allowed by
  the semi-leptonic asymmetry $a_{sl}^d$ is contained within the
  orange boundaries.}
\label{C1}
\end{figure}

\begin{figure}
\begin{center}
\includegraphics[width =0.45 \textwidth]{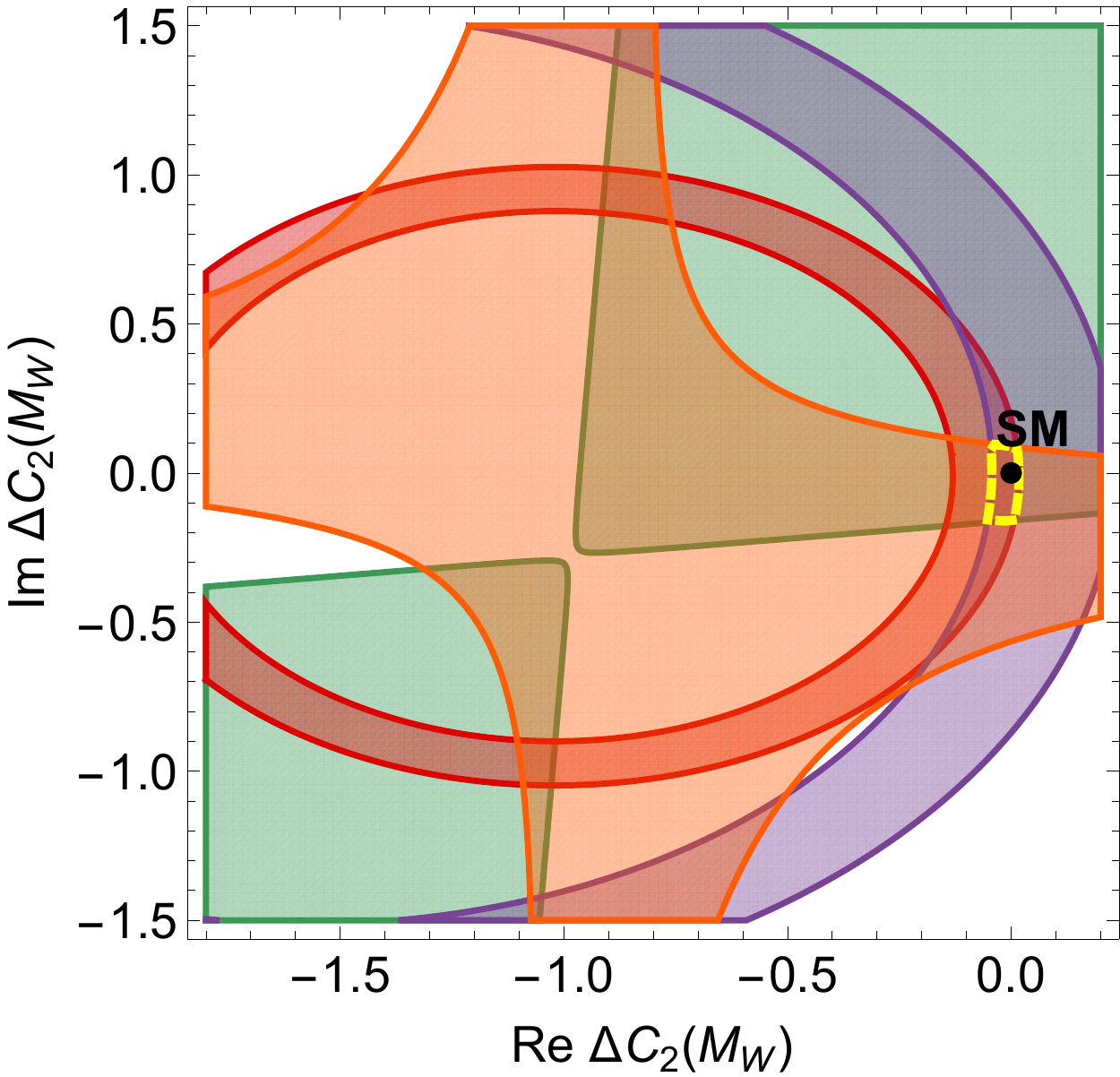}
\end{center}
\caption{Constraints on $\Delta C_2$, the new-physics contribution to
  the tree-level Wilson coefficient $C_2$, at the scale $\mu_W =
  M_W$. The red and purple rings enclose the bounds from the decays $B
  \to D \pi$ and $B \to X_s \gamma$, respectively. The orange
  star-shaped region is related to the semi-leptonic asymmetry
  $a_{sl}^d$. The constraint from $B \to \pi \pi$ comes from the
  observable $S_{\pi\pi}$ and is visualised by the green sections.}
\label{C2}
\end{figure}

To obtain the constraints on new-physics contributions to $C_1$ and
$C_2$ we perform a parameter scan for all the observables described
above, combining all errors in quadrature. In Fig.~\ref{C1} and
Fig.~\ref{C2} we show the regions allowed by each observable at 90\% CL;
for clarity we restrict ourselves to the observables that lead to the
strongest bounds. Moreover, we did not consider possible cancellations
among the new contributions to $C_1$ and $C_2$, i.e. when investigating the
bounds on $\Delta C_1(M_W)$, we set $\Delta C_2(M_W) = 0$ and vice
versa.


We read from the plot the following ranges as rough estimates for
possible new-physics contributions to the current-current operators:
\begin{eqnarray}
{\rm Im} \, \Delta C_1 \in  [-0.56 ;+0.13] \; ,  && 
{\rm Im} \, \Delta C_2 \in  [-0.17;+0.10] \; ,
\label{bounds1}
\\
{\rm Re} \, \Delta C_1 \in  [-0.17 ;+0.12] \; ,  && 
{\rm Re} \, \Delta C_2 \in  [-0.06;+0.02] \; .
\label{bounds2}
\end{eqnarray}
More quantitative statements will be obtained in~\cite{next}. Note
that the bounds obtained in~\cite{Bobeth:2014rra} from $B \to K^{(*)}
\pi/\rho$ observables would slightly shrink the regions given in
Eq.~(\ref{bounds1}) and Eq.~(\ref{bounds2}), but this does not change
our main conclusion: that new physics effects in ${\rm Im} \, C_1$,
${\rm Re} \, C_1$, and ${\rm Im} \, C_2$ can easily be of order
$10\%$.

\section{Precision in $\gamma$}\label{sec:gamma}

We will now study the implications of our findings for the expected
precision of the extraction of the CKM angle $\gamma$ from tree-level
decays. It is defined by $\gamma \equiv \arg(-V_{ud}^{\phantom{*}}
V_{ub}^*/V_{cd}^{\phantom{*}} V_{cb}^*)$ and can be determined from
$B^\pm \to D K^\pm$ decays that receive contributions only from
tree-level operators~\cite{Bigi:1981qs}. The fact that all relevant
hadronic matrix elements can be obtained from data and the absence of
penguin contributions leads to the exceptional theoretical cleanness
of this determination.

The sensitivity to the angle $\gamma$ arises via the interference
between the $b \to c \bar{u} s$ and the $b \to u \bar{c} s$ decay
amplitudes. Denoting the $B^- \to D K^-$-amplitude by $A_1 e^{i
  \delta_1}$ and the $B^- \to \bar{D} K^-$-amplitude by $A_2 e^{i
  (\delta_2 - \gamma)}$, where we have made the dependence on the CKM
angle $\gamma$ explicit, we get
\begin{eqnarray}
{\cal{A}} (B^- \to f_D  K^-) & = & A_1 e^{i \delta_1} \left[ 1 + r_B e^{i (\delta_B - \gamma)} \right] \; ,
\\
{\cal{A}} (B^+ \to f_D  K^+) & = & A_1 e^{i \delta_1} \left[ 1 + r_B e^{i (\delta_B + \gamma)} \right] \; ,
\end{eqnarray}
with $r_B = A_2/A_1$ and the difference of the strong phases $\delta_B
= \delta_2-\delta_1$. The interference of the two decay modes is
achieved via common final states $f_D$ of the decaying $D^0$ and
$\bar{D}^0$ mesons. Different methods to extract $\gamma$ have been
devised, conventionally distinguished according to the different $D$
decay modes. In the GLW method~\cite{Gronau:1990ra,Gronau:1991dp} one
uses $D$ decays into CP eigenstates. In the ADS
method~\cite{Atwood:1996ci,Atwood:2000ck} a combination of
Cabibbo-favoured and doubly Cabibbo-suppressed $D$-decays is chosen
such that interference effects are maximised. Finally, in the GGSZ
method~\cite{Giri:2003ty} three-body $D$ decays are studied with a
Dalitz-plot analysis. Subsequently, further methods were studied, see
e.g. the review in~\cite{Antonelli:2009ws}.

The angle $\gamma$ has been measured by BaBar~\cite{Lees:2013nha} and
Belle~\cite{Poluektov:2010wz, Aihara:2012aw}. Currently the best
experimental precision is achieved by the LHCb collaboration which
quotes $\gamma = \big(73^{+9}_{-10}\big)^\circ$~\cite{LHCbckm2014} for
their ``robust'' combination which includes only $B \to D K$
modes. However, the $B \to D \pi$ modes where the smaller interference
term is compensated by larger branching ratios also start to play a
role in the extraction of $\gamma$~\cite{LHCbckm2014}.

Theoretical corrections to the extraction of $\gamma$ were
investigated extensively in the literature. The effects of $D -\bar D$
mixing and of CP violation in $D$ and also $K$ decays (for final
states with neutral kaons) have been studied in~\cite{Silva:1999bd,
  Grossman:2005rp, Bondar:2010qs, Rama:2013voa, Martone:2012nj,
  Grossman:2013woa}. These effects lead to shifts in $\gamma$ of at
most a few degrees and can be taken into account exactly by a suitable
modification of the expressions for the amplitudes. The shifts can be
larger in the $B \to D \pi$ modes. The irreducible theoretical
uncertainty is due to higher-order electroweak corrections and has
been found to be negligible for the extraction of $\gamma$ using the
$B \to D K$ modes~\cite{Brod:2013sga}. It is expected to be tiny also
in the $B \to D \pi$ case~\cite{Brod:2014}.

Given the expected sensitivity of order $1^\circ$ at
LHCb~\cite{CERN-LHCC-2011-001} and Belle II~\cite{Abe:2010gxa} we now
address the following question: How large of a shift in $\gamma$ due
to new-physics contributions in tree-level decays is still allowed by
data?
In order to compute the shift in $\gamma$ induced by $\Delta C_1$ and
$\Delta C_2$ we start from the effective Hamiltonians for $b \to c
\bar{u} s$ and $b \to u \bar{c} s$ decays. We will consider the two
amplitudes
\begin{equation}
A (B^- \to D^0 K^-) = \langle D^0 K^-| {\cal H}_{\rm eff.}^{\bar{c}us}|B^-\rangle
\; \; \;
\mbox{and}
\; \; \;
A (B^- \to \bar{D}^0 K^-) = \langle \bar{D}^0 K^-| {\cal H}_{\rm eff.}^{\bar{u}cs}|B^-\rangle
\; .
\end{equation}
The CKM angle $\gamma$ can be extracted from the ratio of these two
amplitudes via
\begin{equation}
r_B e^{i (\delta_B - \gamma)} = \frac{A (B^- \to \bar{D}^0 K^-)}
                                     {A (B^- \to       D^0 K^-)} \; .
\end{equation}
Inserting the expressions for the effective
Hamiltonian~\eqref{eq:Heff} we get
\begin{equation}
r_B e^{i (\delta_B - \gamma)} = \frac{V_{ub} V_{cs}^*}{V_{cb} V_{us}^*}
                                 \frac{\langle \bar{D}^0 K^-| Q_2^{\bar{u}cs}|B^-\rangle}
                                      {\langle       D^0 K^-| Q_2^{\bar{c}us}|B^-\rangle}
\left[
\frac{C_2 + r_{A'} C_1}{C_2 + r_A C_1}
\right] \; ,
\end{equation}
where we defined the additional amplitude ratios
\begin{equation}\label{eq:rA}
\begin{split}
r_{A'} = \frac{\langle \bar{D}^0 K^-| Q_1^{\bar{u}cs} | B^- \rangle}
              {\langle \bar{D}^0 K^-| Q_2^{\bar{u}cs} | B^- \rangle} \; ,
\quad 
r_A = \frac{\langle       D^0 K^-| Q_1^{\bar{c}us} | B^- \rangle}
              {\langle       D^0 K^-| Q_2^{\bar{c}us} | B^- \rangle} \; .
\end{split}
\end{equation}
Note that here the Wilson coefficients should be evaluated at the
scale $\mu_b \sim m_b$; we assume this convention throughout the
current section. The estimates given in Eq.~\eqref{bounds1}
and~\eqref{bounds2} correspond to the following ranges at scale
$\mu_b$, obtained using RG running at LO:
\begin{eqnarray}
{\rm Im} \, \Delta C_1 \in  [-0.62 ;+0.14] \; ,  && 
{\rm Im} \, \Delta C_2 \in  [-0.19 ;+0.11] \; ,
\label{bounds1b}
\\
{\rm Re} \, \Delta C_1 \in  [-0.19 ;+0.13] \; ,  && 
{\rm Re} \, \Delta C_2 \in  [-0.066 ;+0.022] \; ,
\label{bounds2b}
\end{eqnarray}
New physics effects in $C_1$ and $C_2$ then modify the ratio $r_B e^{i
  (\delta_B - \gamma)}$ as
\begin{equation}
r_B e^{i (\delta_B - \gamma)} \to
r_B e^{i (\delta_B - \gamma)} \cdot
\left[
\frac{C_2 + \Delta C_2 + r_{A'} ( C_1 + \Delta C_1)}{C_2 + r_{A'} C_1}
\frac{C_2 + r_A C_1}{C_2 + \Delta C_2 + r_A (C_1 +\Delta C_1)}
\right] \; .
\label{exact}
\end{equation}
Thus any new complex contribution to $C_1$ and/or $C_2$ will introduce
a shift in $\gamma$.
Using that $|C_1/C_2| \approx 0.22$ at the scale $m_b$ and that also
$|\Delta C_1/C_2|$ and $|\Delta C_2/ C_2|$ are small 
(see Sec.~\ref{sec:NP}) we can further simplify the above 
relation by expanding in these small ratios:
\begin{equation}
r_B e^{i (\delta_B - \gamma)} \to
r_B e^{i (\delta_B - \gamma)} \cdot
\left[ 1+ (r_{A'}-r_A) \frac{\Delta C_1}{C_2} \right] \; ,
\end{equation}
which depends now only on the modification of the Wilson coefficient
$\Delta C_1$. This modification leads then to a modified value of
$\gamma$
\begin{equation}
\gamma \to \gamma + \delta \gamma = \gamma + (r_A - r_{A'}) \frac{{\rm Im} 
\Delta C_1}{C_2}. 
\end{equation}
Here the dominant dependence of the shift in $\gamma$ on ${\rm Im}
\Delta C_1$ can be nicely seen; for numerical evaluations we
recommend, however, to use the exact expression in Eq.(\ref{exact}).

In order to relate the bounds in Eq.~\eqref{bounds1b}
and~\eqref{bounds2b} to the shift in $\gamma$ we need to estimate the
ratios of matrix elements~\eqref{eq:rA}. Naive colour counting and
neglecting the annihilation topology in $r_{A'}$ gives $r_{A} \approx
{\cal O} (1)$ and $r_{A'} \approx {\cal O} (N_c)$, where $N_c=3$ is the
number of colours. On the other hand, naive factorisation yields
\begin{equation}
r_A     \approx \frac{f_D F_0^{B \to K}(0)}{f_K F_0^{B \to D}(0)}
    \approx 0.4 \; ,
\end{equation}
whereas including the annihilation topology would reduce $r_{A'}$. There
are certainly large uncertainties on these estimates, but it seems
very unlikely that the two ratios cancel accidentally. As a
conservative estimate we will take $r_A - r_{A'} \approx -0.6$.
Having ${\rm Im} \Delta C_{1}(m_b)$ of order $\pm 0.1$ we get $\delta
\gamma$ of order $\mp 4^\circ$, with large uncertainties due to the
hadronic matrix elements.

\section{Conclusion and Outlook}\label{sec:conclusion}
We have investigated constraints on new physics contributions to the
tree-level Wilson coefficients $C_1$ and $C_2$, arising from a set of
observables in the $B$-meson sector. We find that sizable deviations
from the SM are still possible. Specifically, we find that the allowed
ranges of Re$\Delta C_1$, Im$\Delta C_1$ and Im$\Delta C_2$ are of the
order of $10\%$, whereas the allowed range for Re$\Delta C_2$ is slightly
smaller.

A new-physics contribution to the imaginary parts of $C_1$ and $C_2$
plays a particularly important role in view of the precise
determination of the CKM angle $\gamma$ from tree-level decays. The
possible presence of a new weak phase in $C_1$ and $C_2$ introduces an
uncertainty into the extraction of $\gamma$, the latter essentially
being defined as the phase of the CKM element $V_{ub}^*$. The ranges
given in Eq.~\eqref{bounds1} and~\eqref{bounds2} induce an uncertainty
of $|\delta \gamma| \approx 4^\circ$ which is not negligible in view
of the expected sensitivity of $1^\circ$ at LHCb and Belle~II.

To reduce this uncertainty the bounds on $\Delta C_1$ and $\Delta C_2$
should be improved. For instance, the bound on $\Delta C_1$ depends
sensitively on the semi-leptonic asymmetry $a_\text{sl}^d$. For
instance, assuming a decrease of the experimental error for
$a_\text{sl}^{d}$ by 20\% would cut out most of the allowed region for
the imaginary part of $\Delta C_1$ given in Fig.~\ref{C1}. Moreover,
further improvements (both in experiment and theory) in the
observables $R_{D\pi}$, $R_{\rho\rho}$, $R_{\pi \pi}$ and $S_{Dh}$, as
well as an improvement in the theory expression for the total life
time -- e.g. NNLO QCD corrections to the inclusive non-leptonic decay
rates -- would also reduce the allowed parameter ranges for new
physics effects in tree-level decays. We have also seen that the
effect of new weak phases in $C_1$ and $C_2$ on the determination of
$\gamma$ depends sensitively on two ratios of hadronic matrix elements
which are hard to evaluate numerically, and it would be worthwhile to
go beyond our very naive estimates.

Finally, it is worth noting that, conversely, given an independent
measurement of $\gamma$, the CP asymmetries in $B \to D K$ decays
might yield the strongest bounds on new weak phases in the
current-current sector.

In this letter we have attempted only a rough estimate of the new
physics contribution to the tree-level Wilson coefficients; our main
conclusion is that sizable effect cannot be excluded from the
viewpoint of data. Our analysis can be improved in many ways.
First of all, the combination of the different observables was done at
the level of a simple parameter scan, i.e. by computing the 90\% CL
region for each observable separately and intersecting these
regions. Statistical and systematic errors for each observable were
combined in quadrature. For a complete (frequentist) statistical
analysis all observables have to be combined in a single likelihood
function and systematic errors have to be treated within the Rfit
scheme~\cite{Hocker:2001xe}. The combination into a single likelihood
function necessarily reduces the allowed region, but the treatment of
systematic errors in the Rfit scheme typically overcompensates this
effect. In any case, these modifications do not change the result by
orders of magnitude and will therefore have no impact on the main
message of this paper that new-physics effects in $C_1$ and $C_2$ of
the order of 10\% are not in contradiction to data. We postpone a
systematic fit to a future publication~\cite{next} where we will also
investigate flavour specific bounds. More generally, an advanced study
should also allow for new physics contributions to operators other
than exclusively $Q_1$ and $Q_2$.

\section*{Acknowledgements}   
G.TX acknowledges the financial support of CONACyT (Mexico).

\end{document}